\documentclass[pdflatex,sn-mathphys-num]{sn-jnl}
\usepackage{graphicx}
\usepackage{latexsym}
\usepackage{amsthm}
\usepackage{xcolor}
\usepackage{natbib}
\usepackage{subcaption}
\usepackage{textcomp}
\usepackage{xcolor}
\usepackage{url}
\usepackage{multicol}
\usepackage{mathtools}
\usepackage{booktabs} % For professional looking tables
\usepackage{adjustbox}
\usepackage{tcolorbox}
\usepackage{makecell}
\usepackage{tabularx} % 自動改行機能を使用するためのパッケージ
\usepackage{array}    % 表のカラムタイプを定義するのに便利
\usepackage{hhline}

\usepackage{xcolor}
\usepackage{listings}

\definecolor{codegreen}{rgb}{0,0.6,0}
\definecolor{codegray}{rgb}{0.5,0.5,0.5}
\definecolor{codepurple}{rgb}{0.58,0,0.82}
\definecolor{backcolour}{rgb}{0.95,0.95,0.92}

\lstdefinelanguage{ControllerSpec}{
  keywords={set, controllerSpec, controller, safety, liveness, assumption, controllable},
  keywordstyle=\color{blue}\bfseries,
  ndkeywords={idle, cook, moveToBelt, NotOverHeated, MOVE_TO_BELT, NOT_COOKING, CERAMIC, Controllable, G1, C},
  ndkeywordstyle=\color{codepurple},
  comment=[l]{//},
  commentstyle=\color{codegreen}\itshape,
  morestring=[b]",
  stringstyle=\color{red},
}

\lstdefinestyle{mystyle}{
    backgroundcolor=\color{backcolour},
    commentstyle=\color{codegreen}\itshape,
    keywordstyle=\color{blue}\bfseries,
    numberstyle=\tiny\color{codegray},
    stringstyle=\color{red},
    basicstyle=\ttfamily\footnotesize,
    breakatwhitespace=false,
    breaklines=true,
    captionpos=b,
    keepspaces=true,
    numbers=left,
    numbersep=5pt,
    showspaces=false,
    showstringspaces=false,
    showtabs=false,
    tabsize=2,
    language=ControllerSpec % 使用新定义的语言
}

\usepackage{diagbox}

\newtheorem{definition}{Definition}

\renewcommand\proofname{\bf Proof}

\theoremstyle{thmstyleone}

\theoremstyle{thmstyletwo}%

\theoremstyle{thmstylethree}%

\title{Automatic Syntax Error Repair for Discrete Controller Synthesis using Large Language Model}

\author{
Yusei Ishimizu$^{1}$,
Takuto Yamauchi$^{2}$,
Sinan Chen$^{3}$,
Jinyu Cai$^{2}$,
Jialong Li$^{1,2}$\thanks{Corresponding author: lijialong@fuji.waseda.jp},
Kenji Tei$^{1}$
}

\date{}

\def\theaffiliations{
$^{1}$ Institute of Science Tokyo, Tokyo 152-8550, Japan \\
$^{2}$ Waseda University, Tokyo 169-8050, Japan \\
$^{3}$ Kobe University, Kobe 657-8501, Japan
}

\begin{document}

\maketitle

\begin{center}
\theaffiliations
\end{center}

% ---------- abstract ----------
\begin{abstract}

Discrete Controller Synthesis (DCS) is a powerful formal method for automatically generating specifications of discrete event systems. However, its practical adoption is often hindered by the highly specialized nature of formal models written in languages such as FSP and FLTL. In practice, syntax errors in modeling frequently become an important bottleneck for developers—not only disrupting the workflow and reducing productivity, but also diverting attention from higher-level semantic design. To this end, this paper presents an automated approach that leverages Large Language Models (LLMs) to repair syntax errors in DCS models using a well-designed, knowledge-informed prompting strategy. Specifically, the prompting is derived from a systematic empirical study of common error patterns, identified through expert interviews and student workshops. It equips the LLM with DCS-specific domain knowledge, including formal grammar rules and illustrative examples, to guide accurate corrections. To evaluate our method, we constructed a new benchmark by systematically injecting realistic syntax errors into validated DCS models. The quantitative evaluation demonstrates the high effectiveness of the proposed approach in terms of repair accuracy and its practical utility regarding time, achieving a speedup of 3.46 times compared to human developers.
The experimental replication suite, including the benchmark and prompts, is available at \url{https://github.com/Uuusay1432/DCSModelRepair.git}
\end{abstract}


%1
\section{Introduction}
\label{sec: introduction}
In software development, it is crucial to accurately define operational specifications that guarantee safety during the early development stages. In traditional development processes, developers manually formulate and design the operational specifications; and safety is verified through testing after implementation based on these specifications. 
However, this test-based approach not only consumes resources to increase test coverage but also potentially leads to waste in the process (e.g., errors causing rework and redesigning of specifications).
Addressing this challenge, Discrete Controller Synthesis (DCS) has gained attention as a formal method that automatically synthesizes controllers that satisfy requirements such as safety and reachability under anticipated environments. This provides a correct-by-construction approach to designing software specifications, aiming to reduce verification burdens and human errors in later stages.
Recently, DCS has been increasingly applied in various domains, including service mobile robots \cite{10336205}, unmanned aerial vehicle (UAV) missions \cite{10.1145/3513091}, smart homes \cite{Li2024Discrete}, and business process management \cite{BPM}.

Despite its promise, the practical application of DCS hinges on developers’ ability to construct precise and correct formal models of the system’s environment and requirements, often using specialized languages such as Finite State Processes (FSP) and Linear Temporal Logic (LTL). However, this modeling process poses a significant barrier; it is notoriously error-prone, requiring specialized expertise in discrete mathematics and meticulous attention to modeling details. Among the various types of errors, syntax errors—ranging from simple typos to violations of formal grammar—are a frequent and frustrating bottleneck. Although current compilers can catch some of these errors, identifying and fixing them in large, complex models is often time-consuming and non-trivial, disrupting the development workflow and distracting developers from focusing on the more critical task of designing the model's semantics.

Traditionally, developers rely on manually debugging syntax issues guided by compiler error messages. However, this approach is inefficient and does not scale well with increasing model complexity. While rule-based and template-based repair methods—such as dictionary-based spell checkers and abstract syntax tree (AST)-based syntax rule analysis—are commonly used in integrated development environments (IDEs) for popular languages like C or Java, these methods often incur high development and maintenance costs, which are especially burdensome for niche formal methods. Moreover, AST-based techniques struggle to determine the correct insertion points for missing tokens in the text. Recent advances in automated program repair have shown promising results for general-purpose programming languages \cite{Gupta2017DeepFix, Yasunaga2020DrRepair}. However, these approaches are typically data-intensive and not directly applicable to domain-specific modeling languages such as FSP and FLTL, which lack large-scale code corpora. More recently, Large Language Models (LLMs) have demonstrated impressive capabilities in program repair tasks \cite{Liu2023CrashBugs}, yet their potential in the specialized context of formal modeling for DCS remains largely underexplored. Since DCS itself is a niche modeling language, naive zero-shot prompts often fail to enable LLMs to specify DCS-specific grammatical dependencies and structural consistency, resulting in limitations on repair accuracy.

To this end, this paper introduces an automated LLM approach for repairing syntax errors in DCS models through DCS-specific knowledge-informed prompt engineering. Instead of relying on generic prompts, we first conduct a systematic empirical study, involving expert interviews and hands-on student workshops, to identify the patterns and root causes of common syntax errors in FSP/FLTL. We then leverage these findings to design highly effective prompts that equip the LLM with crucial context, including the erroneous code, compiler feedback, formal grammar rules, and illustrative examples of similar errors, thereby guiding it toward accurate and reliable corrections.

The contributions of this study are as follows:
\begin{itemize}
    \item We introduce an empirically derived taxonomy of common syntactic error patterns in DCS modeling, through expert interviews and student workshops. This taxonomy establishes a strong foundation to inform the design of knowledge-informed LLM prompts and enables future extension.
    \item We design a knowledge-intensive prompting strategy informed by our empirical taxonomy, which integrates formal grammar rules, reference models, and typical error examples to guide LLM toward precise repairs of DCS models.
    \item We construct a benchmark for objectively evaluating the performance of automated repair methods by systematically injecting realistic errors, derived from our empirical study, into multiple validated DCS models.
    \item We conduct a quantitative evaluation using our constructed benchmark, which validates the high effectiveness of our proposed method in terms of repair accuracy and demonstrates its practical utility regarding time and cost.
\end{itemize}

The remainder of this paper is organized as follows. Section~\ref{sec:background} provides the necessary background on DCS and its modeling languages. Section~\ref{sec:proposal1} details our empirical investigation into syntax errors, presenting key findings from expert interviews and student workshops. Based on these insights, Section~\ref{sec:proposal2} introduces our proposed LLM-based syntax repair method and its knowledge-informed prompt design. Section~\ref{sec:benchmark} describes the construction of our evaluation benchmark from real-world error cases. Section~\ref{sec:evaluation} presents the comprehensive experimental evaluation of our approach. Finally, we discuss related work in Section~\ref{sec:related} and conclude the paper with an outline of future work in Section~\ref{sec:conclusion}.

%2
\section{Background}
\label{sec:background}
This section aims to provide the foundational knowledge necessary to understand the subsequent parts of this paper. It is structured in two parts: first, we introduce the core concepts of DCS to establish the problem domain of our research. Second, we detail the grammar of FSP and FLTL, the modeling languages used in this study.

\subsection{Discrete Controller Synthesis}
\label{sec:dcs}
DCS is a formal method that automatically synthesizes a software controller to enforce a set of desired properties within a given operational environment~\cite{Ramadge1987}. The core principle of DCS is modeled as a two-player game between the \emph{controller} and the \emph{environment}. The controller aims to select from a set of controllable actions to "win" the game by satisfying the formal requirements, regardless of the uncontrollable actions taken by the environment.

DCS takes two primary inputs: an \emph{Environment Model}, which describes the system's operational context, and a \emph{Requirement Model}, which specifies the desired goals. The Environment Model defines all possible behaviors of the system and its environment, capturing both controllable and monitorable actions. This model is formalized as a Labelled Transition System (LTS). The Requirement Model specifies the properties that the synthesized controller must guarantee. These properties are expressed using Fluent Linear Temporal Logic (FLTL)~\cite{Giannakopoulou2003}, an extension of LTL with the concept of fluents to reason about state changes in event-based systems. FLTL is expressive enough to capture a wide range of temporal properties, including safety ("something bad never happens"), liveness ("something good eventually happens"), and complex assumption-guarantee patterns such as Generative Reactivity.
The output of DCS is a \emph{Controller}, representing a winning strategy in the game. The controller is also an LTS that prunes the behavior of the environment model by disabling certain controllable actions. This ensures the controlled system satisfies all specified properties while remaining non-blocking.

DCS approaches can be broadly categorized based on how they construct and solve the game space. The traditional method, known as full game space construction, first composes the environment and requirement models to build the entire game space~\cite{Magee2006}. The synthesis algorithm then solves the corresponding game (e.g., a reachability game for safety or a Büchi/GR(1) game for liveness) over this complete state space. While this method produces a maximally permissive controller, it suffers from the state-space explosion problem, which limits its scalability.
To address this, advanced techniques such as Directed Controller Synthesis use exploratory or on-the-fly synthesis~\cite{Ciolek2016,Delgado2023}. Instead of constructing the entire state space upfront, these methods begin from the initial state and incrementally explore only the relevant paths that satisfy the requirements. This approach is more scalable but typically supports a narrower range of properties.

\subsection{Grammar of FSP and FLTL in MTSA}
\label{sec: grammar}
This section introduces the grammar used in MTSA (Modal Transition System Analyzer) \cite{mtsa}, a tool for controller synthesis that extends the LTSA (Labeled Transition System Analyzer) tool. In essence, MTSA uses an extension of the Finite State Process (FSP) language to model system behaviors and FLTL to specify system requirements.

Finite State Process (FSP) is used to describe the behavior of concurrent processes for the environment models. An FSP model consists of a set of process definitions, which are composed in parallel to form a complete system model. A process is defined by its states and the actions that cause transitions between them. The fundamental FSP operators include action prefix, choice, and parallel composition. The action prefix operator, denoted by \texttt{->}, specifies sequential behavior. For example, \texttt{CLOCK = (tick -> CLOCK).} describes a process that repeatedly performs the \texttt{tick} action. To introduce non-determinism, the choice operator, \texttt{|}, represents a selection between two or more action sequences, as in \texttt{CERAMIC = (idle -> CERAMIC | cook -> COOKING).}, which can perform either \texttt{idle} or \texttt{cook}. For concurrency, the parallel composition operator, \texttt{||}, composes processes to run simultaneously. Actions with the same name are considered shared and must be executed in synchrony, as in \texttt{||SYS = (PROCESS\_A || PROCESS\_B).}. Further control is provided by guarded actions, using the \texttt{when} keyword to make an action available only if a boolean condition is true. To manage complexity and reuse, FSP supports process labeling, re-labeling, and hiding. Process labeling, using a colon \texttt{:}, prefixes all actions within a process instance (e.g., \texttt{a:SWITCH}), allowing multiple instances to be distinguished. The re-labeling operator, \texttt{/}, changes the names of actions to match interfaces, such as in \texttt{CLIENT/\{call/request, reply/wait\}}. Finally, the hiding operator, \texttt{\textbackslash}, conceals specified actions from the external interface, making them internal \texttt{tau} actions.

FLTL \cite{Giannakopoulou2003} is used in MTSA to define system requirements, which extends standard LTL with the concept of fluents. 
\begin{definition}[Fluent Linear Temporal Logic (FLTL)]
An FLTL formula $\phi$ is defined by the grammar:
\[ \phi ::= \neg \phi \mid \phi \vee \psi \mid X\phi \mid \phi U \psi \mid fl \]
where $\neg$ (negation) and $\vee$ (disjunction) are standard logical operators, $X$ (next) and $U$ (until) are temporal operators, and $fl$ is a fluent. Common syntactic sugar includes $\wedge$ (conjunction), $\rightarrow$ (implication), $\Box$ (always), and $\Diamond$ (eventually).
A fluent $fl$ is defined as $\langle I_{fl}, T_{fl} \rangle_{init}$, where $I_{fl} \subseteq L$ is the set of initiating actions, $T_{fl} \subseteq L$ is the set of terminating actions, with $I_{fl} \cap T_{fl} = \emptyset$, and the subscript $init \in \{\text{true}, \text{false}\}$ indicates the initial value. A fluent becomes true upon the occurrence of an action in $I_{fl}$ and false upon an action in $T_{fl}$.
\end{definition}

For Controller Synthesis Constructs, MTSA introduces specific keywords to define and synthesize a controller. The \texttt{controller} keyword initiates the synthesis process, taking an environment model and a goal specification as input. This goal is defined within a \texttt{controllerSpec} block, which encapsulates the various requirements for the controller. 
The example in Listing~\ref{lst:controller_spec_example} illustrates this structure.

\begin{lstlisting}[caption={Example of a controller synthesis definition.}, style=mystyle,label={lst:controller_spec_example}]
// Define the set of controllable actions
set Controllable = {idle, cook, moveToBelt}.

// Define controller goals
controllerSpec G1 = {
  safety = {NotOverHeated},
  liveness = {MOVE_TO_BELT},
  assumption = {NOT_COOKING},
  controllable = {Controllable}
}.

// Synthesize the controller
controller ||C = (CERAMIC)~{G1}.
\end{lstlisting}

\section{Domain Knowledge Acquisition through Expert Interviews and Student Workshops}
\label{sec:proposal1}

To gain a deep understanding of the current landscape and specific categories of syntax errors in DCS, we adopted a two-pronged approach combining expert interviews with observational student workshops. This investigation was not only intended to capture empirical evidence of common error patterns, but also to establish a structured taxonomy that would serve as a foundation for our LLM-based repair framework. Such an operational taxonomy is crucial for enabling effective in-context learning (ICL) \cite{dong2024surveyincontextlearning},  as representative examples organized by error category allow the LLM to correctly recognize DCS-specific syntactic error patterns and perform high-precision repairs. First, we conducted semi-structured interviews with five researchers experienced in DCS to obtain qualitative insights into typical error patterns. Second, we organized a DCS workshop followed by a hands-on development session, in which student participants engaged in actual modeling tasks. 

\subsection{Semi-Structured Expert Interviews}

We conducted semi-structured interviews with five domain experts in DCS. These experts had over two years of research experience with DCS and included two university faculty members with doctoral degrees, one Ph.D. student, and two master's students. The interviews were conducted individually in a semi-structured format, allowing for both guided discussion and open-ended elaboration. All interviews were recorded with consent and subsequently transcribed for thematic analysis. The resulting qualitative data informed both the categorization of errors and the design of subsequent user studies.

Each interview was guided by a predefined set of open-ended questions designed to elicit expert perspectives on common modeling errors. Key areas of inquiry included the perceived frequency and nature of syntax and semantic errors, the situational contexts in which these errors are likely to occur, and the underlying cognitive or technical challenges that contribute to such mistakes. Participants were also asked to reflect on particularly challenging aspects of DCS model construction, such as the complexity of the formal notation, difficulties in debugging, or expressing modeling intent. Additionally, they were invited to share concrete examples from their past modeling experiences, including how specific errors were identified and eventually resolved.

\subsection{Model Development Workshops}

To complement the qualitative findings from expert interviews and to obtain empirical evidence of error patterns in DCS modeling, we conducted a controlled workshop involving actual model development tasks. The goal was to systematically capture the types, frequencies, and resolution processes of syntax-related errors encountered by participants during hands-on development.

The workshop involved five student participants, including two master's students and three undergraduates. None of the participants had prior experience with DCS modeling. To prepare them for the task, we conducted a short introductory tutorial covering the fundamental concepts of DCS, the syntax of LTS, and the use of the modeling environment. This tutorial was delivered at the beginning of the workshop and included basic examples to ensure baseline familiarity.

Each participant was then given a modeling task derived from a realistic automated warehouse system specification and asked to construct a DCS model accordingly \cite{hope_for_the_best}. The task was designed to reflect common modeling challenges while remaining accessible to beginners. The workshop was conducted over four sessions, each lasting two hours, totaling approximately eight hours of hands-on modeling time per participant. All modeling activities were performed on the participants’ own laptops in a uniform development environment configured with Git version control.

With the participants’ informed consent, the entire modeling process was automatically and rigorously recorded using Git. For each commit, we collected (i) the complete code snapshot of the LTS model, (ii) the compilation status (success or failure), and (iii) any compiler error messages. This setup enabled detailed reconstruction of the modeling trajectory, including the introduction and resolution of specific errors. The collected data were analyzed through automated log processing and code differencing. Specifically, we identified the commits that failed to compile and compared them with their closest preceding or succeeding successful commits to isolate the changes responsible for the errors. By correlating these changes with the associated error messages, we classified the type of error and identified its precise location in the code. This approach allowed for the mechanical identification of common error categories such as typographical mistakes, malformed expressions, and incorrect syntax usage.

\subsection{Key Findings}
\begin{figure}[htbp]
    \includegraphics[width=1\linewidth]{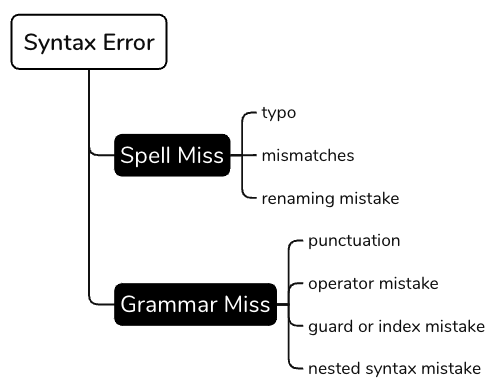}
    \caption{Key Findings through Expert Interviews and Student Workshops.} 
    \label{Summary_findings}
\end{figure}

Through expert interviews and student workshops, we have constructed systematic criteria to categorize the error patterns, as summarized in Fig.~\ref{Summary_findings}.

Broadly speaking, syntax errors in DCS modeling fall into two main categories: spelling mistakes and grammar mistakes. 
For spelling mistakes, which include typos, inconsistent use of letter case, and naming mismatches across different parts of the model—for example, failing to update all references after renaming an identifier. While such mistakes may not always trigger compiler errors directly, they can lead to unintended behavior or semantic confusion in downstream development.
Another major subtype is grammar mistakes, which involve violations of the formal grammar rules of FSP/FLTL. These include missing or incorrect punctuation (such as commas or periods), misplacement of logical or temporal operators, mismatched parentheses, and misdescriptions in more complex constructs such as guarded actions or indexed structures.

The primary causes of these syntax errors are ordinary human mistakes—such as typing errors—as well as the inherent difficulty of maintaining structural and naming consistency across large or intricate models. While compiler-generated error messages offer immediate feedback and are useful for flagging potential issues, identifying the precise location of the error and applying the correct fix becomes increasingly time-consuming and labor-intensive as model complexity grows.

It should be noted that the error taxonomy presented in Fig.~\ref{Summary_findings} is based on the typical patterns observed at the current stage of our empirical study, and therefore does not claim to cover all possible syntax errors in DCS modeling. As the future technical evolution and practical application of DCS, new error types that fall outside the present categories may emerge. To address this, it is possible to continuously refine and evolutionarily extend the taxonomy, incorporating newly observed patterns into the framework, thereby maintaining its utility for guiding repair models.

Additionally, while the error categories in Fig. \ref{Summary_findings} appear superficially similar to those in general-purpose programming languages like C or Java \cite{Gupta2017DeepFix}, the existing tools developed for syntax error repair cannot be directly applied to DCS.
First, unlike widely used languages, FSP is based on process algebra, where punctuation determines the structural composition of the system. A simple punctuation error can fundamentally alter the system state-space structure, requiring a repair mechanism that understands the rigorous semantics of formal methods.
Second, widely used languages benefit from mature compilers that provide precise localization and actionable error messages to facilitate syntax repair. In contrast, niche formal modeling tools like MTSA lack the capability to precisely localize syntax errors. The logs output by the system often produce opaque or misleading error messages, making both automated repair and manual debugging disproportionately difficult.

% \textbf{Semantic errors} are logical errors where the model itself compiles successfully, but its behavior deviates from the developer's intent. Unlike syntax errors, these are difficult for compilers to detect, and their discovery at later stages can significantly increase correction costs.
% \begin{itemize}
% \item Intended transitions are not accurately represented.
% \item Desired safety properties are not satisfied.
% \item The monitoring model itself contains modeling errors.
% \item Outputs result in toCompose is Null, indicating potentially unachievable requirements.
% \item Challenges related to distinguishing between controllable and uncontrollable events are also included.
% \end{itemize}
 % For semantic errors, the main challenges lie in accurately formalizing complex safety requirements into a model and the tendency to overlook scenarios that lead to unintended behaviors. Debugging these errors is a highly complex and demanding task, often requiring extensive effort to identify unintended paths within vast state spaces.

\section{Automatic Syntax Correction}
\label{sec:proposal2}

Building upon the insights from our empirical study in the previous section, this section proposes an LLM-based correction tool to address the challenges of syntax errors in DCS model development, aiming to enhance development efficiency.

\subsection{Overview Procedure}
Given the input of a DCS model, it first checks whether the model compiles successfully. If compilation fails, and if the error was a syntax error, the LLM is employed to correct spelling and grammatical mistakes. Specifically, for spelling correction, the LLM leverages general knowledge and code context to detect and fix likely misspellings. For grammar correction, the compiler error message is passed to the LLM, which rewrites the model to resolve the issue. This process yields a model that compiles successfully.

% \begin{algorithm}[thb!]
%     \caption{Procedure of Syntax Correction via LLM and MTSA}
%     \label{alg:llm_mtsa_loop_alg2e}
%     \KwIn{Initial erroneous model $M_{ini}$}
%     \KwOut{Corrected model $M_{crt}$}

%     \BlankLine

%     $i \leftarrow 0, M_0 \leftarrow M_{ini}$\;

%     $(CmplSucc_{i}, ErrorMsg) \leftarrow \text{MTSA}(M_i)$\;

%     \If{$CmplSucc_{i}$ \text{== TRUE}}{
%         \Return{$M_i$}\;
%     }

%     $M_i \leftarrow \text{LLM}(M_i,  \text{Context})$ \tcp{Correct Spell Error}

%     \While{true}{
%         $(CmplSucc_{i}, ErrorMsg) \leftarrow \text{MTSA}(M_i)$\;
%         \If{$CmplSucc_{i}$ \text{is SUCCESS}}{
%             \Return{$M_i$}\;
%         }
%         $i \leftarrow i + 1$\;
%         $M_{c_{i}} \leftarrow \text{LLM}(M_i,  \text{Context}, \text{ErrorMsg})$ \tcp{Correct Grammar Error}
%     }

%     \Return{$M_i$}\;

% \end{algorithm}

\subsection{Prompt Design}
This section outlines the prompt design process in our proposed method. The prompts are structured in JSON format to provide a clear and hierarchical representation of information for the LLM. Due to space limitations, we provide only a brief introduction. Interested readers can refer to the GitHub repository for the detailed prompt.

The first part is the "Domain Knowledge Base", which provides the LLM with specialized background knowledge on DCS modeling. This primarily includes two parts. First, the Overview of DCS and Development Procedure offers definitions of DES safety, the overall DES development process, the definition and components of DCS, and standard procedures for model synthesis using DCS. This information enables the LLM to understand the role of the model being corrected within the broader DCS development lifecycle. Second, the FSP/FLTL Grammar Rules provide formal specifications of the grammar used in FSP (for environment modeling) and FLTL (for requirement modeling). Each grammar rule is assigned a unique identifier, allowing the LLM to reference specific rules in its reasoning and explanations. This grammar knowledge forms the foundation for the LLMs to assess and enhance the syntactic correctness of models.

The second part consists of reference model examples, which support In-Context Learning (ICL) by helping the LLM develop a comprehensive understanding of the overall structure and detailed elements of DCS models. This example model is an art gallery security system~\cite{10.1145/3194133.3194146} that operates to control the number of people in a room by allowing/denying entry and includes sections for the environment model, monitoring model (requirements), and controller. Through exposure to these reference examples, the LLM learns the functions and interrelationships among different model sections, as well as common modeling patterns. 

The third part is a curated set of error correction examples, also aimed at enabling ICL. These examples teach the LLM to recognize specific error patterns and apply appropriate correction strategies. Rather than being arbitrarily selected, they are drawn directly from the empirical analysis presented in Section \ref{sec:proposal1}, and thus reflect the most frequent and representative mistakes encountered by developers.

The fourth part is the task instruction, which provides the LLM with the specific instruction and dynamic input relevant to the current correction task. This includes three elements: (i) the task instructions clearly specify whether the LLM is expected to perform spelling correction or grammar correction; (ii) for grammar correction tasks, the compiler-generated error message is included; and (iii) the full LTS or FLTL model is to be corrected.

\section{Benchmark Construction}
\label{sec:benchmark}

This section details the construction of a syntax error benchmark dataset based on MTSA models and error injection informed by empirical analysis.

\subsection{Base Model Selection}
For the base models, we selected four MTSA models: Air Traffic (AT), Bidding Workflow (BW), Cat and Mouse (CM), and Access Management (AM). These models have been widely used in prior studies~\cite{Ciolek2016, Delgado2023} and are publicly available on the MTSA tool's website\footnote{\url{https://mtsa.dc.uba.ar/}}. In addition, they exhibit various features of DES, such as parallel composition, non-deterministic choice, guarded transitions, and indexed processes.

Specifically, the AT model simulates aircraft landing coordination to avoid collisions on ramps and at holding altitudes. The BW model captures project evaluation in an engineering firm, where documents are either approved or discarded based on team reviews. The CM model involves guiding mice through a corridor while avoiding cats, requiring strategic movement. The AM model represents a privacy-aware bus streaming system with access control, synchronization, and resource constraints.

\subsection{Syntax Error Injection Method}

The second step is syntax error injection. This process aims to replicate realistic error scenarios identified in our investigation.

For spelling mistake injection, we drew from the findings in Section~\ref{sec:proposal1}, which identified "typing errors" and "inconsistencies across multiple occurrences" as the primary causes of spelling mistakes. To mimic simple typing errors noted by experts, we injected misspellings at random locations by applying common patterns such as adjacent key substitutions, character duplications, and omissions. To reproduce inconsistencies across multiple occurrences, as observed in participant experiments, we selectively injected spelling mistakes into only a subset of repeated event or process names, leaving the remaining instances unchanged.

For grammar mistake injection, we directly reproduced deviations from LTS notation grammar rules (e.g., missing commas, incorrect operator placement) identified in our earlier investigation. Specifically, we removed essential syntactic elements such as termination symbols (missing periods), action sequence delimiters (missing commas), and choice separators (|), which typically lead to compiler errors like "process identifier expected" or "dot expected." We also introduced operator misuse and incorrect placement, such as misusing assignment operators (= and -), parentheses (\{\} and ()), and parallel composition symbols (||), which can cause messages like "= expected" during compilation. In addition, we injected errors related to index and guard expressions that stem from the complexity of LTS grammar, including incorrect period counts in range definitions, confusion between [i] and .i in indexed actions, and malformed when-clause parentheses.

\subsection{Benchmark Dataset Composition}
Finally, to construct a more comprehensive benchmark dataset, we systematically generated multiple versions of error-injected models based on each selected base model. 
For each of the four base models (AT, BW, CM, AM) selected in Step 1, we created a fixed set of parameterized instances reflecting different levels of complexity and scale within their respective domains. A key selection criterion was the computational feasibility on a standard personal computer, ensuring that all model instances could be parsed and analyzed within a reasonable timeframe(no longer than one minute). 
Specifically, two distinct instances were generated for each base model by varying their inherent parameters, resulting in a total of eight unique model instances. Syntax errors were then injected into each of these instances using the methods described in Step 2.
The quantity of injected errors was defined as a fixed number per instance, covering both spelling and grammatical error types as categorized in Step 2. Care was taken to ensure that each major section of a model (e.g., process definitions, requirements, controller specifications) had a high likelihood of containing at least one syntax error.

As a result, we constructed eight evaluation benchmark models, collectively containing 32 spelling errors and 39 grammatical errors, for a total of 71 injected syntax errors. Each model is guaranteed to include at least one syntax error. Furthermore, careful human check was performed to ensure that the injected errors did not become semantic errors that would alter the model's intended behavior.

\section{Evaluation}
\label{sec:evaluation}

This section is dedicated to the empirical evaluation of our proposed syntax error correction tool, using the benchmark developed in the previous section. The evaluation aims to address the following research questions.

\begin{itemize}
\item \textbf{RQ1: Error Identification and Correction}
How accurately can the tool identify and correct the location of injected syntax errors?

\item \textbf{RQ2: Computation Overhead and Practicability}
How practical is the tool in terms of correction time, monetary cost, and number of iterations required until successful correction?
\end{itemize}

\subsection{Experimental Setup}

We conducted the experiment using the benchmark models introduced in Section \ref{sec:benchmark}, which contain intentionally injected syntax errors. A model is considered successfully corrected if it compiles without errors. To account for stochastic variability in LLM outputs, we repeated each experiment three times under the same settings and report the average results.

For RQ1, we conduct two types of evaluation: error identification, assessed at the line level, and error correction, which evaluates whether the applied modification successfully fixed the error. If the tool fails to resolve a grammatical error within four consecutive correction attempts, the correction is deemed unsuccessful. Each line of the model is categorized as True Positive (TP), False Positive (FP), False Negative (FN), or True Negative (TN), depending on whether the LLM attempted a modification and whether the line actually contained an error. A line is considered a TP if it contains a syntax error and the tool attempts to modify it for identification, or if it contains a syntax error, is modified, and the modification successfully resolves the error for correction. An FP occurs when a syntactically correct line is unnecessarily modified. An FN arises when a line with a syntax error is left unmodified for identification, or when the tool either fails to modify it or modifies it but fails to fix the error for correction. Finally, a TN is recorded when a syntactically correct line is correctly left unchanged.

Based on these classifications, we further compute precision, recall, and F1-score.
As the baseline, we compare our approach, which incorporates instructions, domain knowledge, and examples of models and errors, with a zero-shot approach that only includes instructions, in order to evaluate the effectiveness of the proposed prompt design.

For RQ2, we compare the tool’s usability against human performance by measuring the time required to correct syntax errors. 
To evaluate human performance, we conducted a controlled experiment involving five participants: four graduate students and one senior undergraduate in Computer Science. Four of these participants were actively engaged in DCS-related research, possessing theoretical knowledge and modeling experience ranging from six months to two and a half years. The remaining participant had no direct DCS research experience but had acquired preliminary modeling experience by participating in the workshop described in Section \ref{sec:proposal1}.
A repair was considered complete once the model compiled successfully. Each participant was allotted a maximum of ten minutes per model, with the task terminating upon either successful completion or timeout.
For the LLM performance evaluation, we assessed the tool’s usability by measuring the total correction execution time per model, focusing specifically on the time attributable to the LLM, including prompt construction, API calls, and response parsing. All LLM-based experiments were conducted using MTSA (Modal Transition System Analyzer) \cite{mtsa} on a MacBook equipped with an M4 chip and 16 GB of RAM, employing OpenAI’s GPT-4.1 via the API as the representative LLM.
%経験者-都竹、生方、大畑 not経験者-小山、青柳

\subsection{Experimental Results}

\begin{table*}[t!h]
\centering
\caption{Results of RQ1-1: Error identification, where N refers to the number of injected errors. }
\label{tab:rq1_1_results}
\begin{tabular}{|c|c|c||c|c|c|c|c|c|c|c|}
\hline
\multicolumn{3}{|c|}{}  & \multicolumn{4}{|c|}{0-shot}  & \multicolumn{4}{|c|}{Proposal}\\
\hline
\textbf{Model} & \textbf{Error Type} & \textbf{N} & \textbf{Acc} & \textbf{Pre} & \textbf{Recall} & \textbf{F1} & \textbf{Acc} & \textbf{Pre} & \textbf{Recall} & \textbf{F1}\\
\hline
CM(2,2) & Spelling  & 3 & 99\% & 100\% & 89\% & 0.94 & 100\%  & 100\% & 100\% & 1.0\\
 & Grammar  & 2  & 93\% & 0\% & 0\% & 0.0 & 95\%  & 33\% & 100\% &0.50\\
\hline
CM(3,3) & Spelling  & 6 & 97\% & 81\% & 72\% & 0.76 & 98\%  & 100\% & 72\% & 0.84\\
 & Grammar  & 4 & 95\% & 52\% & 100\% & 0.69 & 97\%  & 67\% & 83\% & 0.74\\
\hline
BW(2,2) & Spelling  & 4 & 92\% & 75\% & 25\% & 0.38 & 93\%  & 100\% & 25\% & 0.4\\
 & Grammar  & 4 & 89\% & 43\% & 50\% & 0.46 & 90\%  & 46\% & 50\% & 0.48\\
\hline
BW(5,2) & Spelling  & 4 & 100\%  & 100\% & 100\% & 1.0   & 100\%  & 100\% & 100\% & 1.0\\
 & Grammar  & 7 & 87\% & 33\% & 33\% & 0.33 & 97\%  & 75\% & 100\% & 0.86\\
\hline
AM(2,5) & Spelling  & 5 & 99\%  & 94\% & 100\% &  0.97 & 100\%  & 100\% & 100\% &  1.0\\
 & Grammar  & 4 &  88\% & 27\% & 50\% & 0.35 & 93\%  & 50\% & 67\% & 0.57\\
\hline
AM(3,5) & Spelling  & 5 & 100\%  & 100\% & 100\% & 1.0  & 99\%  & 94\% & 100\% & 0.97\\
 & Grammar  & 12 & 59\% & 25\% & 64\% & 0.36 & 97\%  & 88\% & 100\% & 0.94\\
\hline
AT(2,2) & Spelling  & 3 & 100\%  & 100\% & 100\% &  1.0 & 100\%  & 100\% & 100\% & 1.0\\
 & Grammar  & 2 & 99\% & 86\% & 100\% & 0.92 & 100\%  & 100\% & 100\% & 1.0\\
\hline
AT(4,4) & Spelling  & 2 & 95\% & 25\% & 67\% & 0.36 & 100\%  & 100\% & 100\% & 1.0\\
 & Grammar  & 4 & 98\% & 69\% & 75\% & 0.72  & 100\%  & 100\% & 100\% & 1.0\\
%\hline
%HC(N,2) & Spelling  & 4 & \% & \% & \\
% & Grammar  & 6 & \% & \% & \\
%\hline
%HC(S,5) & Spelling  & 7 & \% & \% & \\
% & Grammar  & 4 & \% & \% & \\
\hline
Sum & Spelling & 32 &  99\%  & 82\%  & 82\%  & 0.82 & 99\%  & 99\% & 85\% & 0.91\\
 & Grammar & 39 & 90\% & 34\% & 59\%  & 0.43 & 97\%  & 70\% & 90\% & 0.79\\
\hline
\end{tabular}
\end{table*}

\begin{table*}[t!h]
\centering
\caption{Results of RQ1-2: Error correction, where N refers to the number of injected errors.}
\label{tab:rq1_2_results}
\begin{tabular}{|c|c|c||c|c|c|c|c|c|c|c|c|c|c|c|}
\hline
\multicolumn{3}{|c|}{} & \multicolumn{4}{|c|}{0-shot}  & \multicolumn{4}{|c|}{Proposal}\\
\hline
\textbf{Model} & \textbf{Error Type} & \textbf{N} & \textbf{Acc} & \textbf{Pre} & \textbf{Recall} & \textbf{F1} & \textbf{Acc} & \textbf{Pre} & \textbf{Recall} & \textbf{F1}\\
\hline
CM(2,2) & Spelling  & 3  & 99\% & 100\% & 89\% & 0.94 & 100\%  & 100\% & 100\% & 1.0\\
 & Grammar  & 2  & 96\% & \textbf{--}  & 0\%  & 0.0  & 98\%  & \textbf{--} & 0\% & 0.0\\
\hline
CM(3,3) & Spelling  & 6 & 98\% & 100\% & 72\% & 0.84 & 98\%  & 100\% & 72\% & 0.84\\
 & Grammar  & 4 & 97\% & 60\% & 100\% & 0.75 & 98\%  & 83\% & 83\% & 0.83\\
\hline
BW(2,2) & Spelling  & 4 & 93\% & 100\% & 25\% & 0.40 & 93\%  & 100\% & 25\% & 0.4\\
 & Grammar  & 4 & 88\% & 36\% & 33\% & 0.34 & 93\%  & 67\% & 50\% & 0.57\\
\hline
BW(5,2) & Spelling  & 4 & 100\%  & 100\% & 100\% & 1.0  & 100\%  & 100\% & 100\% & 1.0\\
 & Grammar  & 7 & 90\% & 47\% & 33\% & 0.39 & 96\%  & 80\% & 76\% & 0.78\\
\hline
AM(2,5) & Spelling  & 5 & 100\%  & 100\% & 100\% & 1.0  & 100\%  & 100\% & 100\% &  1.0\\
 & Grammar  & 4 & 91\% & 38\% & 50\% & 0.43 & 95\%  & 36\% & 58\% & 0.45\\
\hline
AM(3,5) & Spelling  & 5 & 100\%  & 100\% & 100\% & 1.0  & 100\%  & 100\% & 100\% & 1.0\\
 & Grammar  & 12 & 81\% & 50\% & 64\% & 0.56 & 89\%  & 82\% & 50\% & 0.62\\
\hline
AT(2,2) & Spelling  & 3 & 100\%  & 100\% & 100\% &  1.0 & 100\%  & 100\% & 100\% & 1.0\\
 & Grammar  & 2 & 100\% & 100\% & 100\% & 1.0  & 100\%  & 100\% & 100\% & 1.0\\
\hline
AT(4,4) & Spelling  & 2 & 98\% & 60\% & 50\% & 0.55 & 100\%  & 100\% & 100\% & 1.0\\
 & Grammar  & 4 & 98\%  & 82\% & 75\% & 0.78  & 100\%  & 100\% & 100\% & 1.0\\
\hline
Sum & Spelling & 32 & 99\% & 98\% & 81\% & 0.89 & 99\%  & 100\% & 85\% & 0.91\\
 & Grammar & 39 & 94\% & 52\% & 57\% & 0.54 & 97\%  & 82\% & 64\% &  0.72\\
\hline
\end{tabular}
\end{table*}

\begin{table}[t!hb]
\centering
\caption{Results of RQ2: Tool execution time, token usage, and API cost per model. TT = Average Tool correction Time (sec), HT = Average Human correction Time (sec).}
\label{tab:rq2_results}
\begin{tabular}{|c|c|c|c|c|c|c|}
\hline
\textbf{Model} & \begin{tabular}[c]{@{}c@{}}\textbf{TT} \\ (s)\end{tabular} & \textbf{Cycle} & \begin{tabular}[c]{@{}c@{}}\textbf{HT} \\ (s)\end{tabular} & \begin{tabular}[c]{@{}c@{}}Input  \\ Token\end{tabular} & \begin{tabular}[c]{@{}c@{}}Output  \\ Token\end{tabular} & \begin{tabular}[c]{@{}c@{}}\textbf{Cost}  \\ (USD)\end{tabular}\\
\hline
CM(2,2) & 114 & 6 & \textbf{186} & 29,442 & 8,570 & 0.10 \\
CM(3,3) & 96 & 4 & \textbf{375} & 41,874 & 12,559 & 0.12\\
BW(2,2) & 64 & 5 & \textbf{345} & 22,960 & 2,382 & 0.06\\
BW(5,2) & 93 & 6 & \textbf{348} & 33,371 & 5,758 & 0.09\\
AM(2,5) & 89 & 4 & \textbf{271} & 24,273 & 4,032 & 0.07\\
AM(3,5) & 48 & 5 & \textbf{212} & 29,867 & 4,906 & 0.10\\
AT(2,2) & 59 & 2 & \textbf{262} & 12,125 & 4,710 & 0.06\\
AT(4,4) & 30 & 1 & \textbf{145} & 5,275 & 2,496 & 0.02\\
\hline
Average & 77.4 & 4.1 & 268 & 24,898 & 5,677 & 0.07\\
%59.8
\hline
\end{tabular}
\end{table}

\begin{figure*}[t]
    \centering
  \begin{minipage}[b]{0.49\hsize}
    \centering
    \includegraphics[width=1\linewidth]{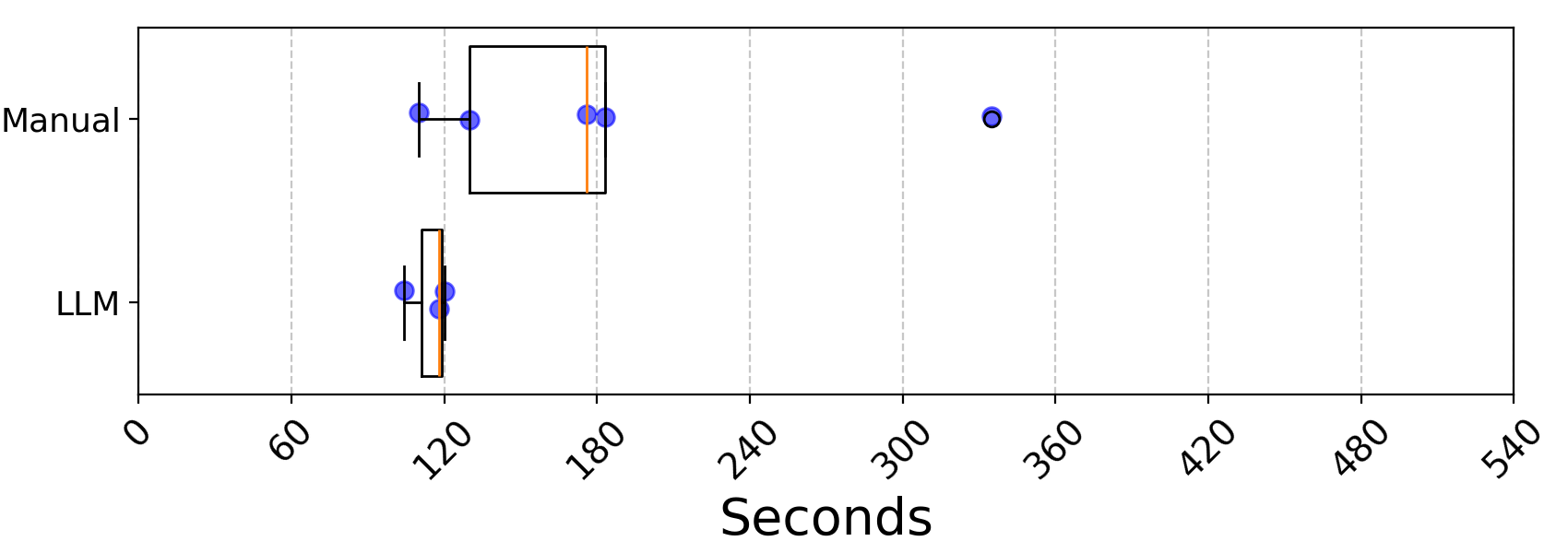}
    \subcaption{CM(2,2)}
  \end{minipage}
  \begin{minipage}[b]{0.49\hsize}
    \centering
    \includegraphics[width=1\linewidth]{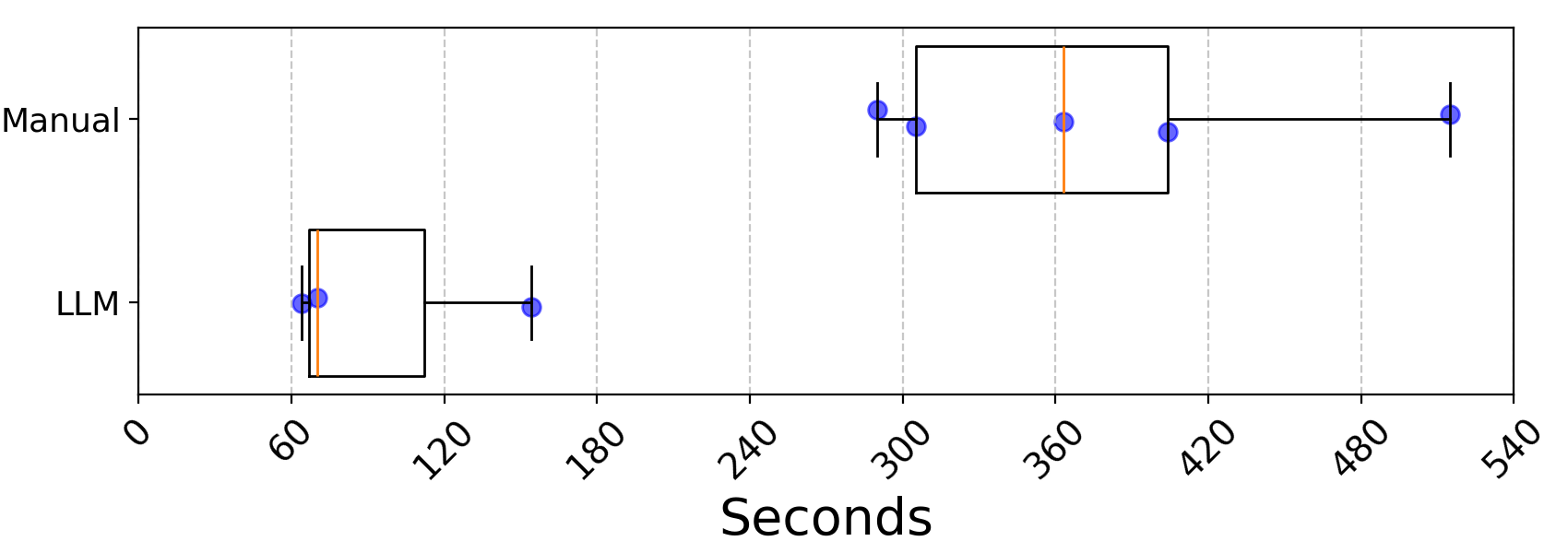}
    \subcaption{CM(3,3)}
  \end{minipage}
  \begin{minipage}[b]{0.49\hsize}
    \centering
    \includegraphics[width=1\linewidth]{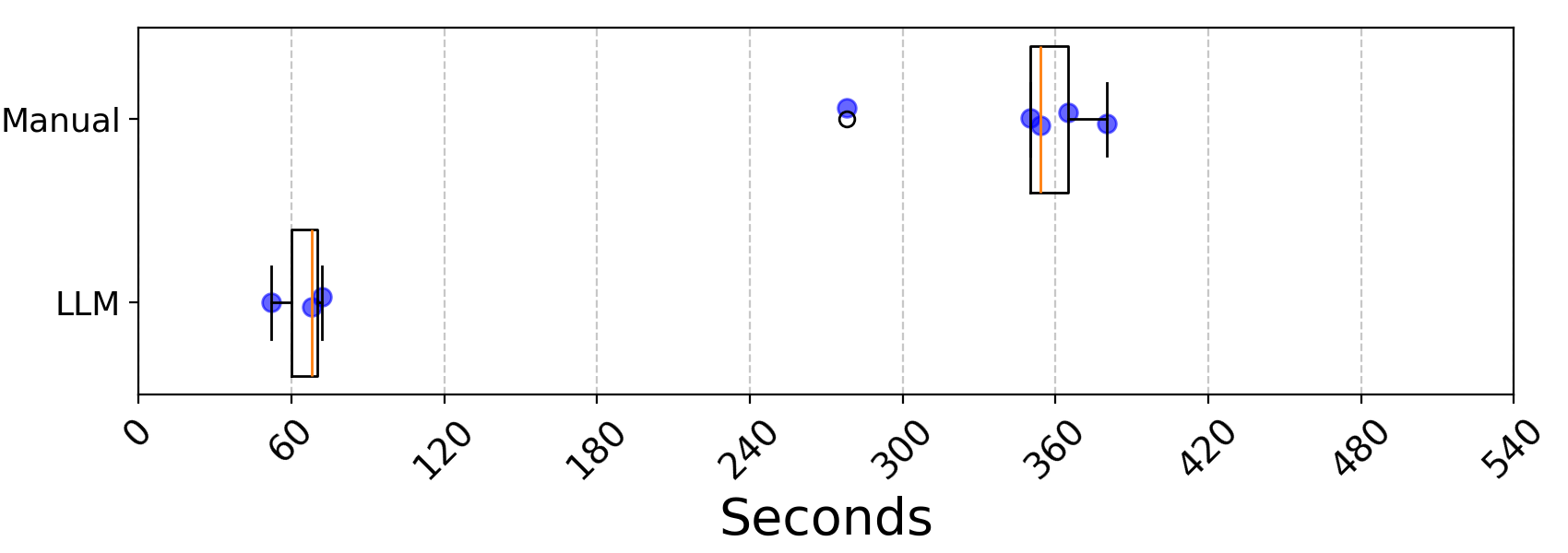}
    \subcaption{BW(2,2)}
  \end{minipage}
    \begin{minipage}[b]{0.49\hsize}
    \centering
    \includegraphics[width=1\linewidth]{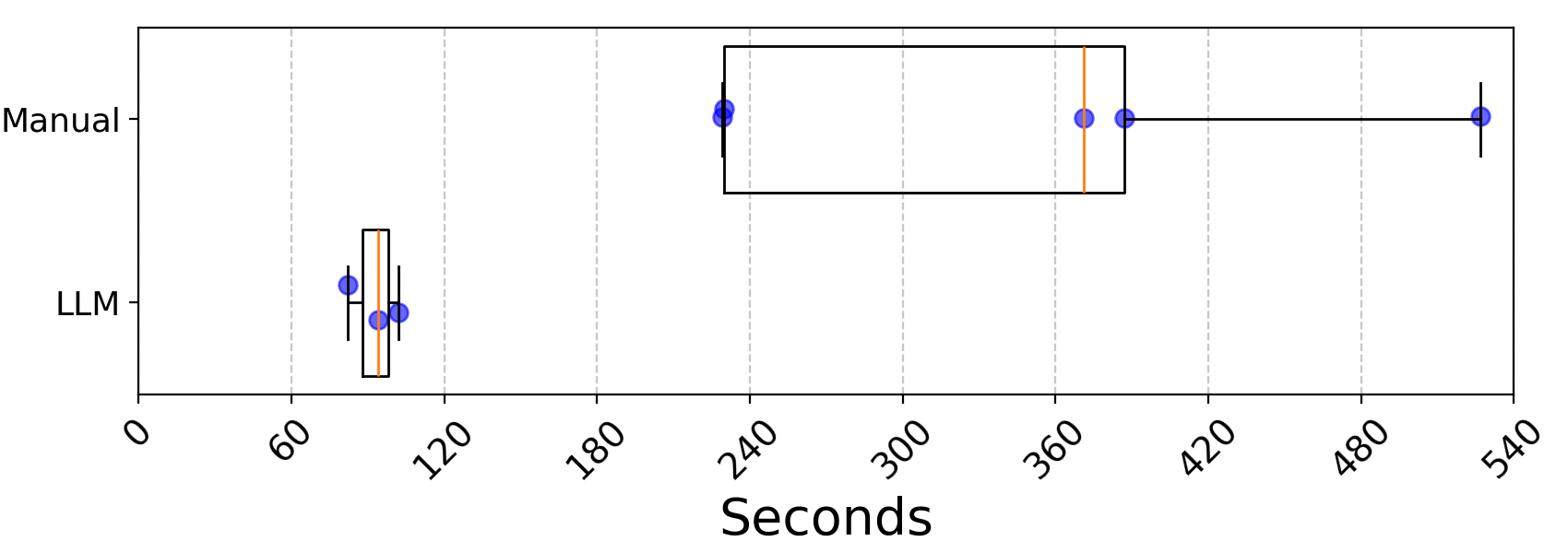}
    \subcaption{BW(5,2)}
  \end{minipage}
  \begin{minipage}[b]{0.49\hsize}
    \centering
    \includegraphics[width=1\linewidth]{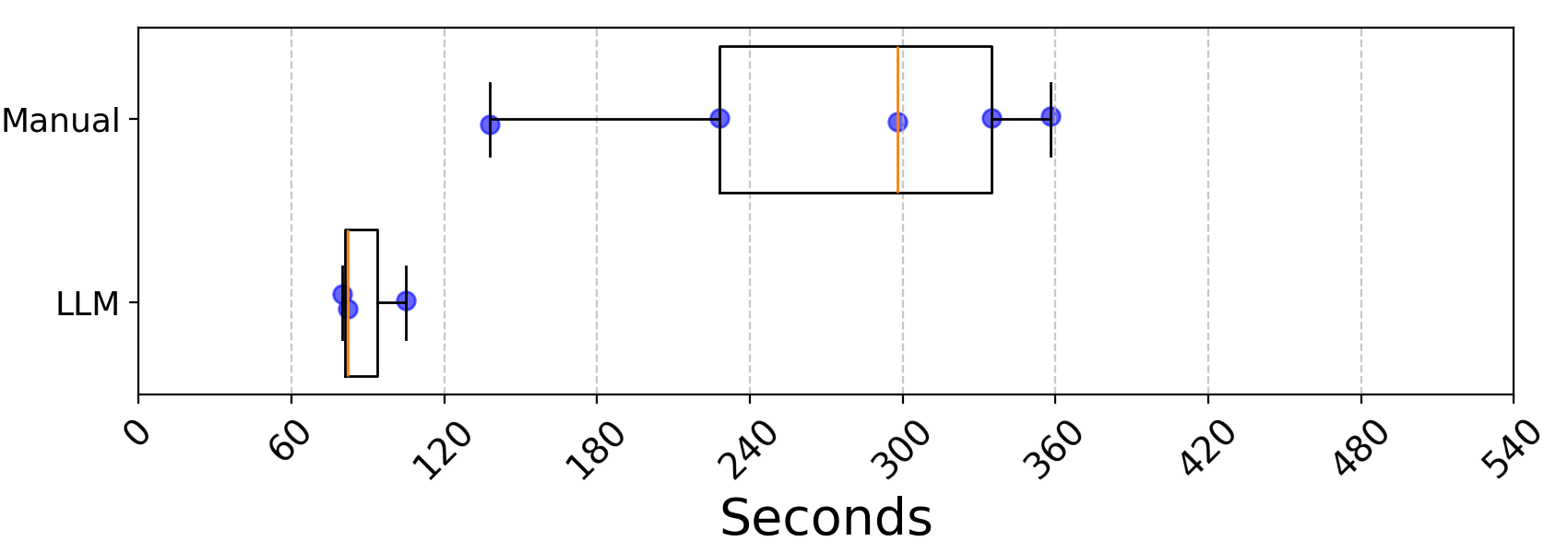}
    \subcaption{AM(2,5)}
  \end{minipage}
  \begin{minipage}[b]{0.49\hsize}
    \centering
    \includegraphics[width=1\linewidth]{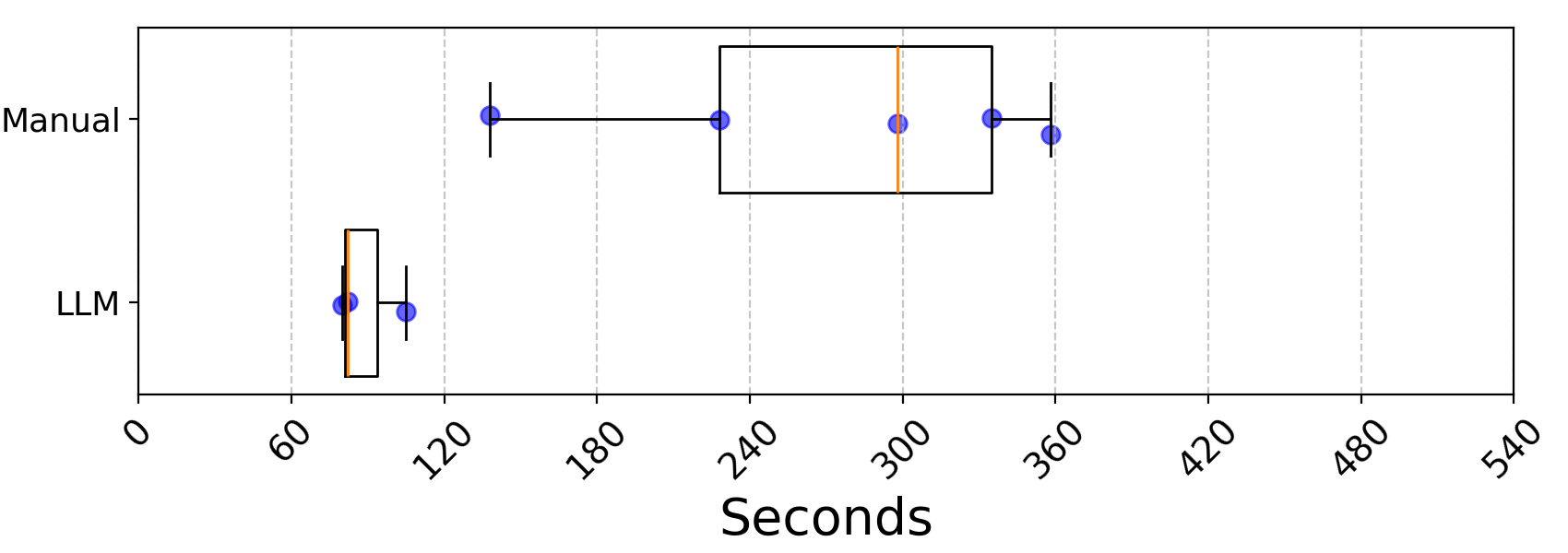}
    \subcaption{AM(3,5)}
  \end{minipage}
  \begin{minipage}[b]{0.49\hsize}
    \centering
    \includegraphics[width=1\linewidth]{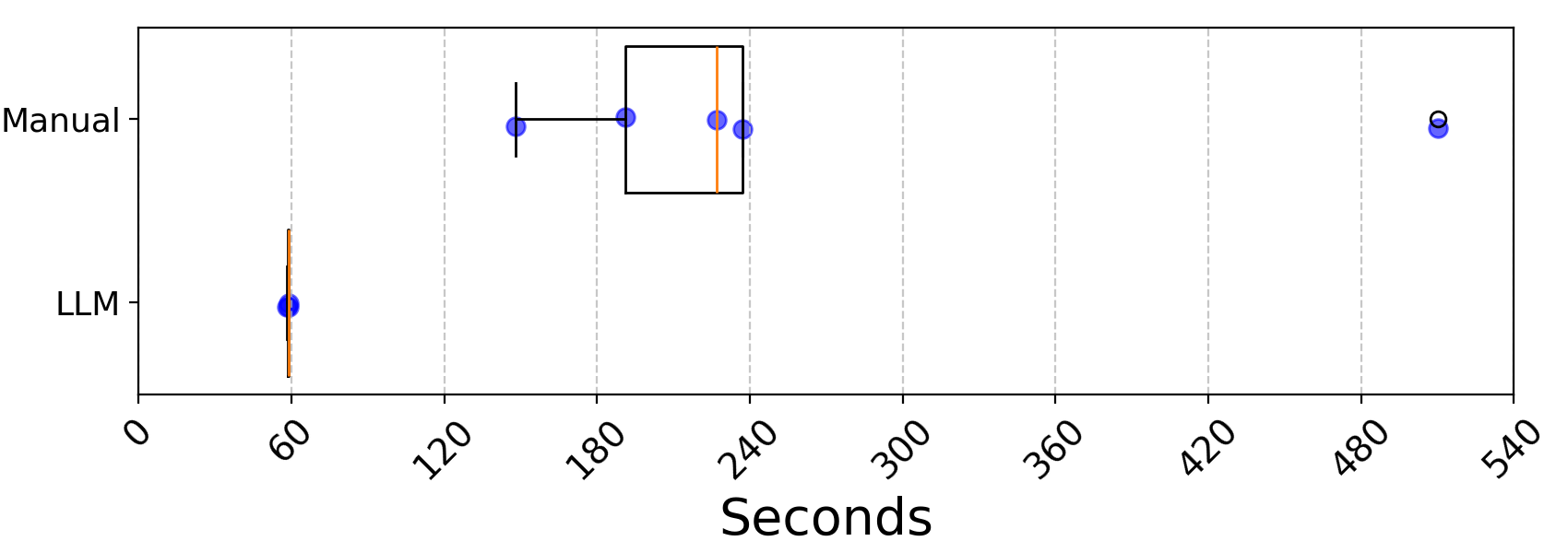}
    \subcaption{AT(2,2)}
  \end{minipage}
    \begin{minipage}[b]{0.49\hsize}
    \centering
    \includegraphics[width=1\linewidth]{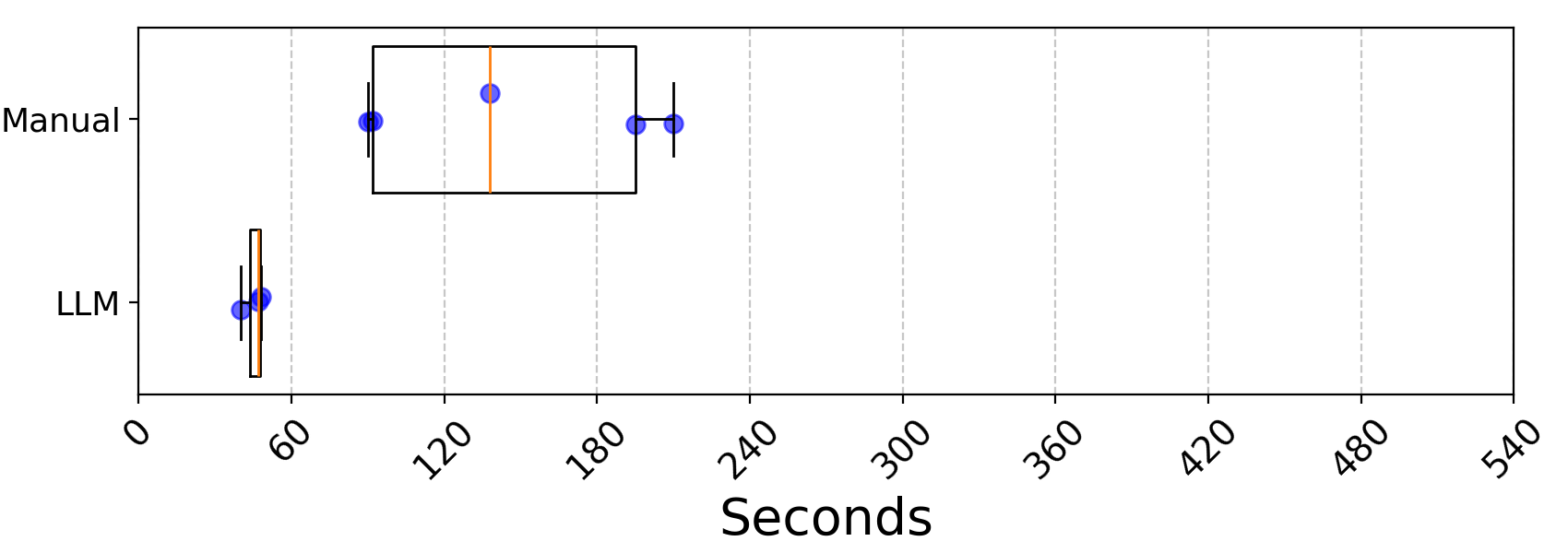}
    \subcaption{AT(4,4)}
  \end{minipage}
  \caption{Distribution of Error Correction Time.}
  \label{fig:distribution}
\end{figure*}

For RQ1, Table~\ref{tab:rq1_1_results} and Table~\ref{tab:rq1_2_results} summarize the results. The key finding is that while both approaches are highly effective for spelling correction, our proposal demonstrates a significant advantage in the more complex task of grammar correction.  
For spelling error identification, our proposal consistently achieved higher scores across all metrics (accuracy 99\%, precision 99\%, recall 85\%, F1-score 0.91), whereas the zero-shot method also performed reasonably well but fell short of perfection (accuracy 99\%, precision 82\%, recall 82\%, F1-score 0.82). This indicates that although zero-shot can identify many spelling errors, our proposal is markedly more reliable.
For spelling error correction, both approaches also recorded strong performance. The zero-shot method reached accuracy 99\%, precision 98\%, recall 81\%, and an F1-score of 0.89, showing that it can correct many spelling mistakes but often introduces unnecessary changes (i.e., modifying words that did not require correction). By contrast, our proposal maintained consistently high scores with fewer such extraneous modifications, highlighting its robustness. Notably, in certain configurations, the difference becomes striking; for example, with the AT(4,4) model, the zero-shot method yielded a relatively low F1-score of 0.55, while our proposal achieved a perfect score of 1.0.
For more complex grammar errors, the advantage of our proposal becomes more evident. The summary data shows a substantial lead in the F1-score (0.72 vs.\ 0.54), driven by a dramatic improvement in precision (82\% vs.\ 52\%) and a notable gain in recall (64\% vs.\ 57\%). This trend is consistent across most models; for example, with the BW(5,2) model, our proposal achieved an F1-score of 0.78, more than doubling the zero-shot method's score of 0.39.

For RQ2, we assess the practical efficiency of our method in terms of execution time, token usage, and API cost, with detailed results presented in Table~\ref{tab:rq2_results}. On average, the tool correction time (TTT) per model was 77.4 seconds. In contrast, human developers required an average of 268.0 seconds (HT), indicating that our tool is approximately 3.46 times faster than manual debugging for the same tasks.
Furthermore, Fig.~\ref{fig:distribution} illustrates the distributional comparison between human and LLM correction times. As shown in the boxplots, human performance exhibits high variance. While human debugging is prone to substantial delays when participants encounter challenging error patterns—with some cases exceeding 500 seconds—the tool demonstrates consistent and predictable execution times. 
This level of manual debugging effort corroborates the authors' own experience\footnote{Three researchers with more than 10-year DCS research experience.}: during the system modeling phase, verifying the behavior of a partially constructed model often necessitates interrupting the workflow to spend several minutes fixing syntax errors. This disruption significantly impairs the overall continuity and cognitive flow of the modeling activity.
Moreover, in this context, the efficiency of our method is further highlighted by its ability to run in the background (i.e., the tool can be invoked while developers continue their modeling activities). Consequently, the measured execution time of 77.4 seconds does not equate to actual blocking wait time for developers, making the tool even more practical.
Finally, the tool demonstrates strong cost-effectiveness. The average API cost is minimal at only \$0.07 per model repaired. In contrast, the human effort of 268.0 seconds corresponds to approximately \$1.48 per model (assuming an average hourly wage of \$20 for software engineers in Japan), which is more than 20 times the cost of the automated approach.

\subsection{Discussion and Limitations}
We observed several recurring phenomena that reveal deeper challenges in the design and evaluation of LLM-based syntax correction tools. These findings suggest that error correction is a complex process shaped by cross-component interactions, structural context, correction stability, and semantic fidelity.

First, we observed notable cross-component interactions between the spelling and grammar correction modules. In several cases, errors originally caused by incorrect identifiers or model-specific keywords were successfully fixed by the grammar correction component, despite not being flagged as spelling mistakes. For example, in the CM(2,2) model, the identifier \texttt{Levels} was mistakenly written as \texttt{Level}. Although \texttt{Level} is a valid English word and passed the spelling checker, it caused a compile-time error due to a mismatch with the model syntax. The grammar module ultimately replaced it with the correct token. This behavior indicates that the boundary between spelling and grammar errors is often fluid, and that treating correction as an integrated process—rather than a pipeline of isolated components—may yield better results in some situations.

Second, we found that correction success was highly sensitive to local code context and error density, which may be related to internal mechanisms of LLMs such as Transformer architecture and next-token prediction. Specifically, even when the same type of syntax error—such as a missing comma or incorrect guard syntax—was injected across different models, fix rates varied widely. This discrepancy can be attributed to structural context: errors that are isolated and surrounded by otherwise well-formed code are more easily identified and repaired; in contrast, errors clustered within dense, complex process definitions tend to confound the model’s repair logic. For example, when multiple commas were omitted across consecutive lines, the tool consistently failed to identify and fix them correctly.

Third, we observed that repeated attempts to correct persistent syntax errors\footnote{
For details, please see our github:\url{https://github.com/Uuusay1432/DCSModelRepair}} often led to overcorrection, where the tool modified already correct parts of the code and inadvertently introduced new mistakes. For example, the correct snippet \texttt{Team(ID=0) = Arrival} was rewritten as \texttt{Team[ID=0] = Arrival}, resulting in syntactically invalid output. In other cases, the tool correctly identified the problematic line but applied an inappropriate fix, leading to semantically incorrect changes. These observations suggest that current settings (prompts and models) may lack sufficient repair stability and conservativeness, and suggest a future direction to enable explicit constraints to prevent excessive or unnecessary edits.

Fourth, potential biases in both the benchmark dataset and the prompt design, each derived from the empirical study described in Section~\ref{sec:proposal1}, also warrant attention. Since DCS is a niche language, and industrial users (including our industry partners) do not publicly release their models due to intellectual property constraints, our empirical observations are limited to experiences from the academic community and student-generated small-scale datasets. Similarly, our experiment benchmark is adapted from the limited publicly available datasets rather than derived from real-world industrial models. As future work, it will be necessary to refine the prompt design and expand the evaluation by incorporating feedback from both academia and industry, such as newly identified error patterns or newly available datasets.

Finally, a critical limitation lies in the distinction between successful compilation and semantic correctness. Specifically, successful compilation does not necessarily indicate that an error has been correctly fixed. The model may produce modifications that eliminate syntax errors but alter the intended behavior. In extreme cases, the model may remove or rewrite entire logic blocks to eliminate errors, thereby masking the original functionality. Since our evaluation primarily considers compilability, the reported success rates may overestimate the tool’s true effectiveness. Future work should incorporate semantically aware repair validation—such as specification conformance or execution-based tests—to better capture the real impact of generated fixes.

\subsection{Threats to Validity}

One primary threat to internal validity stems from the inherent stochasticity of LLM outputs. To mitigate this, we conducted each experiment three times per setting and reported the average results. We acknowledge that larger-scale experiments—including the use of different LLMs and the computation of confidence intervals—could further enhance the reliability of our findings. However, due to the necessity of interacting with the external MTSA tool, fully automated experiments were not feasible, which practically limited the overall experimental scale.

The second threat concerns the expertise of the human participants: they have between six months and five years of experience in the DCS domain, corresponding roughly to junior-level engineers. As a result, more experienced engineers may require less manual time to perform corrections.

The third factor concerns the selection of benchmark models. This study utilizes four well-established MTSA models (AT, BW, CM, and AM), which are structurally sound and have been widely adopted in academic contexts. As a result, the tool’s performance may partially benefit from the high quality of these curated models. Additionally, our evaluation was conducted at the line level, which may obscure fine-grained edit behaviors. For example, a single line may be partially correct or contain multiple sub-errors, which cannot be disambiguated at this level of granularity. Future work could explore token-level or AST-based evaluation for more precise insights.

The main threat to external validity arises from the way errors are injected into the benchmarks. Although this study adopts a systematic error injection strategy that more closely resembles realistic development mistakes compared to random injection, the injected errors are still artificially constructed. Real-world errors may involve intertwined issues, unconventional edge cases, or ambiguous faults that defy simple categorization. 
Additionally, the generalizability of our findings beyond the selected benchmark models remains uncertain, particularly for those developed by individuals with varying levels of expertise or more complex logic.

\section{Related Work}
\label{sec:related}

\subsection{Neural-based Repair for Syntax and Compile Errors}
\label{sec:rw-neural}

Early learning-based approaches train task-specific models on large corpora of \texttt{<code, error, fix>} triples to repair syntax errors or broader compilation errors.
DeepFix first demonstrated an end-to-end seq2seq architecture that directly rewrites erroneous lines in C programs, achieving full repairs for 27\% of student submissions on the DeepFix benchmark~\cite{Gupta2017DeepFix}.
DrRepair enhanced the input with compiler diagnostics, constructed a program–feedback graph, and applied a graph neural network with self-supervised pre-training, increasing the full-repair rate to 68\% on the same benchmark~\cite{Yasunaga2020DrRepair}.
More recently, TransRepair employed a Transformer encoder–decoder and 1.8 million synthetic error samples to handle longer programs and multi-error scenarios, outperforming prior methods on both the DeepFix and SPoC datasets~\cite{Li2022TransRepair}.
Iaso further ensembled several specialized correctors with a neural locator and data augmentation, reaching a 78.8\% full-repair rate—already comparable to much larger foundation models while requiring significantly less compute~\cite{Li2023Iaso}.
However, due to the scarcity of FSP code, we do not adopt this data-hungry approach.

\subsection{Prompt-based Repair with Large Language Models}
\label{sec:rw-llm}

In parallel, the rise of code-centric LLMs has introduced a data-agnostic alternative: errors are fixed via prompt engineering or lightweight post-processing, without any model retraining \cite{10.1145/3686803}.
Interactive tools such as GitHub Copilot have shown that Codex-style models can correct many simple syntax mistakes on the fly. Empirical studies report that advanced prompting techniques, such as role-playing or multi-turn interaction, significantly improve an LLM’s debugging accuracy across multiple programming languages~\cite{Liu2023CrashBugs}.
Dedicated pipelines have also emerged. SynFix combines compiler error categories with a large pre-trained model and formulates repair as a multi-label editing task, improving accuracy particularly for longer Java programs~\cite{Ahmed2021SynFix}.
LlmFix targets code \emph{generated} by LLMs themselves; its three-stage filtering pipeline (for indentation, truncation, and missing imports) improves HumanEval/MBPP pass rates by an average of 7.5\% across 14 models~\cite{Wen2024LlmFix}.

Our work builds on this paradigm, but is specifically optimized for DCS. In particular, we (1) construct domain knowledge and a curated error set through expert interviews and student experiments, and (2) leverage a general-purpose LLM with compiler-aware prompts to automatically repair FSP/FLTL models—filling a notable gap in the syntax repair of formal modelling languages.

\subsection{Semantic Repair for Formal Models}
Although semantic repair is not the primary focus of this paper, it has been extensively studied in the formal methods community. Semantic repair refers to automated techniques that modify an erroneous or obsolete formal model so that the revised version once again satisfies a desired set of semantic properties. Research in this area spans declarative languages, temporal logic specifications, goal models, and adaptive controllers, employing techniques such as constraint solving, logical reasoning, and learning-based methods.

In the context of declarative Alloy models, early generate–validate frameworks apply mutations to suspicious syntax tree nodes and validate each candidate using \textsc{AUnit} tests~\cite{ARepair18}. To address test-suite overfitting, later work performs bounded, exhaustive SAT-based searches guided by \texttt{assert} statements~\cite{BeAFix21}. Beyond Alloy, MaxSAT encodings have been used to synthesize or repair LTL specifications from both positive and negative traces, while respecting user-defined syntactic constraints~\cite{Atlas24}.

At the requirements engineering level, counterexample-guided learning adapts KAOS goal models in response to changes in environmental assumptions, while preserving the structure of the original model as much as possible~\cite{Alrajeh20}. In reactive controller synthesis, online approaches dynamically relax GR(1) assumptions or goals and re-synthesize a controller at runtime~\cite{Buckworth23}. Meanwhile, control-and-discovery loops incrementally learn unknown environment behaviors modeled as labeled transition systems (LTSs), while concurrently enforcing safety goals~\cite{Keegan22}.

\section{Conclusion}
\label{sec:conclusion}
In this paper, we introduced an automated approach that leverages LLMs to repair syntax errors in DCS models. Drawing insights from expert interviews and user studies, we developed a knowledge-informed prompting strategy and constructed a systematic benchmark of realistic errors. Our evaluation demonstrated that our approach is highly effective at identifying and correcting syntax errors, exhibiting a significant advantage over a zero-shot baseline, particularly for complex grammatical issues. 
Furthermore, our method demonstrates superior practical efficiency, achieving a 3.46$\times$ speedup and reducing costs by over 20 times compared to manual debugging.
 
For future work, we plan to extend our approach in three primary directions.
First, to enhance the practical utility of our method, we will integrate it as a feature within the MTSA tool. This will provide developers with a seamless, built-in syntax correction capability, streamlining the modeling workflow.
Second, building on this integration, we aim to promote the adoption of our tool among researchers and industry practitioners in the DCS community. By gathering real-world feedback on its performance and usability, we can iteratively refine the tool to further improve its efficiency and effectiveness in diverse development scenarios.
Third, we will address the more challenging task of semantic error correction, thereby further enhancing the reliability and efficiency of the DCS development process. Unlike syntax errors, semantic errors occur when a model is syntactically correct but fails to reflect the developer's intent. Correcting these errors requires a deeper level of understanding and reasoning about the model's behavior and its alignment with system requirements.

\section*{Acknowledgments}
The research was partially supported by JSPS KAKENHI (grand number 23H03374, and 25K15290).

\bibliography{sn-bibliography}

\end{document}